# Inverting Singlet and Triplet Excited States using Strong Light-Matter Coupling

*Elad Eizner[1,*], Luis A. Martínez-Martínez[2], Joel Yuen-Zhou[2], Stéphane Kéna-Cohen[1,*]*

1. Department of Engineering Physics, École Polytechnique de Montréal, Montréal H3C 3A7, QC, Canada.

2. Department of Chemistry and Biochemistry, University of California San Diego, La Jolla, California 92093, United States.

**ABSTRACT**

In organic microcavities, hybrid light-matter states can form with energies that differ from the bare molecular excitation energies by nearly 1 eV. A timely question, given recent advances in the development of thermally activated delayed fluorescence materials, is whether strong light-matter coupling can be used to invert the ordering of singlet and triplet states and, in addition, enhance reverse intersystem crossing (RISC) rates. Here, we demonstrate a complete inversion of the singlet lower polariton and triplet excited states. We also unambiguously measure the RISC rate in strongly-coupled organic microcavities and find that, regardless of the large energy level shifts, it is unchanged compared to films of the bare molecules. This observation is a consequence of slow RISC to the lower polariton due to the delocalized nature of the state across many molecules and an inability to compete with RISC to the dark exciton reservoir, which occurs at a rate comparable to that in bare molecules.

**Introduction.**

In the molecular orbital picture, when an electron is promoted from the highest occupied molecular orbital (HOMO) to the lowest unoccupied molecular orbital (LUMO), it can form either an electron-hole state (exciton) with overall singlet ($S = 0$) or triplet ($S = 1$) total spin ($S$). Under electrical excitation of organic thin films, where uncorrelated charges are injected, spin statistics dictate that when oppositely charged carriers meet on a single molecule, they will form triplet excitons three times more often than singlet excitons[1]. In contrast, under optical excitation, singlet excitons are preferentially formed because triplet exciton formation is spin-forbidden from the singlet ground state. However, spin mixing due to weak spin-orbit coupling or the hyperfine interaction can lead to a subsequent transition from the singlet to the optically dark triplet state, a process known as intersystem crossing (ISC). One important characteristic of triplets is that their energy is almost always below that of the singlet due to the positive value of the Coulomb exchange integral. The singlet-triplet gap, $\Delta E_{st}$, is typically on the order of $\sim 0.5 - 1$ eV, which makes thermally activated reverse intersystem crossing (RISC) from the triplet to the singlet very inefficient compared to typical non-radiative decay rates[2]. Triplet excitons in organic molecules are thus long-lived with very low radiative efficiencies for transitions to the ground state, a process known as phosphorescence. This has been a source of important challenges for the development of efficient organic light-emitting diodes (OLEDs) and organic lasers.

A common approach to overcome losses due to triplets is the use of molecules containing heavy noble metals that greatly increase phosphorescence rates due to spin-orbit coupling[1]. This approach has been very successful, leading to 100% internal quantum efficiency in OLEDs. However, organometallic complexes have some disadvantages such as their high cost due to the use of metals such as platinum and iridium, limited photostability in the blue and possible toxicity[3–5]. In recent years, there has been a tremendous effort to develop purely organic emitters that can harvest triplets by converting them back to singlets through processes such as triplet-triplet annihilation or thermally activated delayed fluorescence (TADF)[4–6]. In TADF materials, the singlet-triplet energy gap is minimized (< 0.1 eV), typically through donor-acceptor molecular design which reduces the spatial overlap between HOMO and LUMO wavefunctions. This allows for thermally activated RISC of triplets to the singlet state, followed by efficient fluorescence. It has been shown that this approach can lead to the conversion of nearly all triplets into emissive singlet states[7]. One disadvantage of this approach is that RISC rates are typically slow (~$10^6$ s$^{-1}$), which can lead to large build-ups of triplets at high excitation densities and subsequent losses due to triplet-triplet and triplet-charge quenching.

It has been known for some time that optical microcavities can be used to modify the energy of singlet excitons through the formation of system eigenstates called polaritons. An outstanding and timely question is: to what extent can these modified energetics affect triplet dynamics?[8–10] More importantly, in the ultimate case where singlets and triplets are inverted energetically, can RISC be made very efficient to obtain high TADF efficiencies for a broad range of materials and avoid the build-up of large triplet populations?

Polaritons are light-matter eigenstates that form when singlet electronic transitions are strongly coupled to the vacuum electromagnetic field in an optical cavity. This occurs when the light-matter interaction rate is faster than the photon and electronic decay rates in the system. In the last decade, organic polaritons have proved to be a remarkable platform for the demonstration of fascinating non-linear quantum phenomena at room temperature[11–13], the development of novel optoelectronic device applications[14–17], and modification of chemical reactions[18–23]. Polariton states are separated into two types of dispersive modes—each on different sides of the electronic transition—called lower polaritons (LPs) and upper polaritons (UPs). The minimal separation between LP and UP states occurs when the cavity photon energy is resonant with the electronic transition and is termed the vacuum Rabi energy ($\Omega$). As a result, the energetic separation between LPs and the triplet is smaller than the separation between triplets and the peak singlet absorption. For a large number of molecules, $N$, the Rabi energy $\Omega \propto \sqrt{N/V}$, where $f$ is the transition oscillator amplitude and $V$ is the cavity mode volume. Importantly, because of the large number of molecules in a typical cavity as compared to the number of photonic modes supported by the cavity, $N_{ph}$, strong coupling leads to the formation of $N - N_{ph}$ dark states at the same energy as the singlet transition[24], commonly referred to as the exciton reservoir. Even in the presence of disorder, these states are largely molecular (rather than photonic) in character and, because of their large number, have been shown to dominate the dynamics of polariton relaxation[23,25–28].

Experiments on triplet dynamics where the singlet transition is strongly-coupled to a cavity mode were performed as early as 2007. In that case, it was found that resonantly excited polaritons could ISC to the triplet, but that dark states dominated the overall dynamics and no significant changes in rates were observed.[8] Very recently, experiments with much larger Rabi energies observed a modest reduction of phosphorescence decay rate in strongly-coupled optical microcavity, which was attributed to polariton-enhanced RISC[9]. Phosphorescence rates, however, are sensitive to any changes in the optical environment, such as the presence of mirrors, so it is unclear to which extent changes were due to radiative effects or to polaritonic effects.

Here, we demonstrate the formation of polaritons in a material showing TADF, where we are able to convincingly invert the singlet polariton and the triplet state energies. We directly measure triplet dynamics via the RISC rate, which is independent of any modifications in the optical density of states from the presence of the microcavity. Finally, we want to highlight the importance of doing such experiments in vacuum to avoid artificial modifications of triplet lifetimes due to quenching from triplet oxygen, and in the linear excitation regime. In control experiments, we find that simple encapsulation from molecular oxygen due to the presence of microcavity mirrors can lead to large changes in rates.

**Results and Discussion.**

Figure 1a shows a simplified diagram of the electronic energy levels and rate constants for TADF molecules, and for polaritons in this material. When TADF molecules are optically excited, charge transfer singlets ($^1$CT) are rapidly formed[6]. Radiative decay can then occur either directly as prompt fluorescence or through ISC followed by RISC, in a process known as delayed fluorescence. The thermally activated RISC rate shows a Boltzmann dependence[4,5]:

$$k_{RISC} = A\exp(-\frac{E_a}{k_B T}), \quad (1)$$

where $k_B$ is the Boltzmann factor, $T$ is the temperature and $E_a$ is an activation energy. In principle, the activation energy depends both on the singlet-triplet energy gap and the singlet-triplet reorganization energy, $\lambda_T$. In TADF molecules that show efficient RISC, the activation energy is expected to be similar to the singlet-triplet energy gap[4,5]. As shown in recent works[29–31], the mechanism for ISC and RISC in TADF materials involves strong vibronic coupling between local triplet states ($^3$LE) and the triplet charge transfer states ($^3$CT), followed by spin orbit coupling to $^1$CT states.

In a polariton setup, optical excitation at energies above that of the singlet exciton, such as at the UP or at even higher energies, leads to rapid relaxation to the dark states[25,27,32]. As a result, light-emission is mostly due to relaxation from the dark states to the LPs, followed by the radiative decay of LPs. The latter process occurs on a scale given by the cavity lifetime (< 10 fs), which can be considered as instantaneous compared to the other timescales in the system. This combination leads to *prompt* radiative decay upon optical excitation. In materials that show TADF, however, dark states can also undergo ISC, which through RISC can ultimately populate the LP. This process leads to *delayed* fluorescence due to the slow rate typical of RISC. As shown in Fig. 1a, the delayed fluorescence pathway will involve ISC to the triplet, followed by triplet to dark state RISC with a rate of $k_{T \to DS}$ and triplet to LP RISC with a rate of $k_{T \to LP}$. The first rate is essentially unchanged from that in the bare system, while the second rate is unique to the microcavity. If it could be made faster than $k_{T \to DS}$, it would provide the means for accelerating RISC compared to the case of the bare molecule.

The molecule we chose for this study is a trigonal donor-acceptor of 1,3,5-tris(4-(diphenylamino)phenyl)-2,4,6-tricyanobenzene (3DPA3CN)[33]. This molecule has been shown to have a singlet-triplet energy gap of $\Delta E_{st} = 0.1$ eV, and slow ($k_{RISC} \sim 2 \cdot 10^3$ s$^{-1}$), but efficient thermally-activated RISC. In addition, 3DPA3CN shows no phosphorescence even at low temperature. The absorption and photoluminescence (PL) of a neat film of 3DPA3CN is shown in Fig. 2. The absorption has a peak at 425 nm (2.92 eV) due to a CT$^1$ excited state, and the PL is Stokes-shifted by 0.66 eV with a peak at 549 nm (2.26 eV). PL quantum efficiency (PLQE) was measured using an integrated sphere with vacuum-packed samples (see Methods). The neat film showed a PLQE $\Phi = 86 \pm 3\%$, and when co-deposited with 2,2′,2"-(1,3,5-Benzinetriyl)-tris(1-phenyl-1-H-benzimidazole) (TPBi), the PLQE reaches $\Phi = 98 \pm 3\%$ (see Table 1). Three microcavities were fabricated (see structure in Fig. 1b), consisting of either neat 3DPA3CN with a thickness of 70 nm (MC Neat) or co-deposited TPBi-3DPA3CN (55% − 45% by volume) with thicknesses of either 64 nm (MC 1) or 81 nm (MC 2). The active layer was sandwiched between two 10 nm TPBi buffer layers and 100 nm (30 nm) Ag bottom (top) mirrors. The buffer layers were used to minimize any direct quenching from the metal. In addition, control samples consisting of the active and buffer layers without any mirrors were fabricated during the same deposition run by masking the metal deposition.

To extract the polariton dispersion relation, we measured angle-resolved reflectivity of the fabricated microcavities, which is shown for transverse electric (TE) light polarization in Fig. 3a-3c respectively. For each incident angle we observe two minima corresponding to the excitation of the LP and UP states. The dashed blue lines show the uncoupled singlet exciton absorption peak ($E_x$) and the cavity photon energies. The black lines show a least squares fit of the minima to the Hopfield Hamiltonian (see Methods). The solid white line shows the triplet energy, at 2.42 eV. The triplet energy was estimated to be 0.1 eV lower in energy than the *bottom* of the singlet band, as reported previously[33], and in agreement with our measurements. The energy of the bottom of the singlet, which corresponds to the 0-0 transition, was taken to be at 2.52 eV, from the crossing energy between the PL and absorption spectra (see Fig. 2). This should be contrasted with the absorption maximum, which corresponds to the vibronic transition with the strongest Franck-Condon overlap. From the fits to the Hopfield Hamiltonian, for MC 1 (Fig. 3a), we obtain a Rabi energy $\Omega = 0.45 \pm 0.02$ eV with detuning of $\Delta = E_c - E_x = -0.36 \pm 0.02$ eV for the TE mode, where $E_c$ is the cavity photon energy at normal incidence, and $\Omega = 0.40 \pm 0.01$ eV for the transverse magnetic (TM) mode (see Table S1).[34] As can be seen from Fig 3a, the LP energies at all incident angles are higher than the triplet energy for MC 1. For a thicker TADF layer, i.e. MC 2 (Fig. 3b), we obtain a Rabi energy $\Omega = 0.50 \pm 0.02$ eV with detuning of $\Delta = -0.49 \pm 0.03$ eV for the TE mode and $\Omega = 0.42 \pm 0.02$ eV for the TM mode. As can be seen from Fig. 3b, the LP energies for incident angles above 35 degrees are higher than triplet energy, but remarkably, for incident angles below 35 degrees the LP energies are lower than triplet energy. Finally, for a neat TADF layer (MC Neat), we obtain a Rabi energy $\Omega = 0.87 \pm 0.01$ eV with detuning of $\Delta = -0.58 \pm 0.02$ eV for the TE mode and $\Omega = 0.87 \pm 0.02$ eV for the TM mode. This Rabi energy corresponds to 30% of the uncoupled singlet exciton energy ($\Omega/E_x$), entering the realm of ultrastrong light-matter coupling[35]. As can be seen from Fig. 3c, the LP energies for all incident angles are lower than the triplet energy, completely inverting the singlet-triplet ordering of the system for all of the measured angles.

Notably, the measured PLQEs for the microcavities were found to be systematically lower than those of the control samples (see Table 1). The lower QEs can be attributed to slow relaxation from the dark states to the LP, relaxation of dark states to TM metal-insulator-metal plasmonic modes with wavevectors beyond the light line, and ohmic losses of the LP mode. As can be seen from the angle-resolved PL spectra in Fig. S1, the PL of all microcavities is dominated by LP emission. The lowest PLQE, $\Phi = 7.5 \pm 3\%$, was measured for MC 1 in which a large portion of the LP states have energies that are higher than the energy of bottom of singlet dark states. Indeed, the inability to populate such LP states can be seen in the spectra as a sharp drop in PL intensity for any wavelengths below the 0-0 transition at ~490 nm (see Fig S1a,). Other microcavities, which had energies below the dark states, do not show this effect and had PLQEs of $\Phi = 32 \pm 3\%$ and $\Phi = 20 \pm 3\%$ for MC 2 and MC Neat respectively.

A direct way to examine the effect of triplet transitions to the LP is by measuring delayed fluorescence and comparing the RISC rates of the microcavities to the control samples. The RISC rate can be calculated from experimental observables using the following equation[4,7,36]

$$k_{RISC} = \frac{k_p k_d}{k_{ISC}} \frac{\Phi_d}{\Phi_p}, \quad (2)$$

where $k_p$ and $k_d$ are the prompt and delayed fluorescence rate constants, respectively, $\Phi_p$ and $\Phi_d$ are the PLQYs of the prompt and delayed components ($\Phi = \Phi_p + \Phi_d$), and $k_{ISC}$ is the ISC rate constant. The latter was estimated from the control sample using $k_{ISC} \approx k_p(1 - \Phi_p)$, and assuming $k_{ISC}$ is the dominant non-radiative decay pathway for the singlet.

To obtain RISC rates, transient PL was measured for all of the samples (see Methods). All measurements were performed under vacuum and at laser powers that ensured that the dynamics stayed in the linear regime. The normalized time-dependent PL intensities are shown in Fig. 4a-4c. The large peak close to 0 ms corresponds to the prompt component, followed by the slow delayed fluorescence due to RISC and re-emission. Polariton transients were measured at normal incidence at the corresponding LP energy. To measure the delayed luminescence rates, $k_d$, mean lifetimes were calculated from a multi-exponential fit to the delayed decay luminescence data (see Methods), the values are shown in Table 1. The delayed and prompt efficiencies, $\Phi_d/\Phi_p$, can be calculated by integrating over the PL intensities of the respective delayed and prompt components[4]. As can be seen from Table 1, only slight changes were found between the microcavities and the control samples. For the neat samples, $k_d$ is higher and $\Phi_d/\Phi_p$ is lower compared to TPBi mixed samples. This is due to concentration quenching at the higher molecular concentration, which is known to have a detrimental effect on triplet lifetimes[4].

The prompt PL decays are shown in Fig. S2 and the prompt rates $k_p$, are summarized in Table 1. The prompt lifetimes were found to be $\tau_p \sim 5$ ns, four orders of magnitude shorter than the delayed lifetimes. We find a slightly faster decay rate in the microcavities compared to the controls (up to a factor of ~1.6). This is principally attributed to a modification of the local density of optical states inside the microcavities. However, relaxation from the dark states to the LP can also potentially reduce the prompt lifetime if it is fast enough to compete with the radiative decay rate. Finally, the RISC rates can be calculated using Eq. 2. As can be seen in Table 1, no significant changes are found for any microcavity compared to the corresponding controls.

In addition to room temperature studies, we performed transient PL measurements at lower temperatures as shown in Fig 4b. By reducing the temperature, we increase the effective activation barrier for RISC, $\Delta E_a/k_B T$, making the process less efficient. In principle, this could allow for the appearance of a slower competing rate to $k_{T \to DS}$, which is not observable at room-temperature. As can be seen from Fig. 4b and the values in Table 1, at lower temperatures $\Phi_d/\Phi_p$ and $k_{RISC}$ decrease significantly. Nevertheless, we did not observe substantial modifications in lifetimes or efficiencies in the microcavity compared to the control.

Figure 4d shows the spectra of delayed and prompt components of MC 2 and the control sample. As was observed previously in some TADFs, the delayed spectra of the control is redshifted compared to the prompt spectra[4,7]. The spectral shift was suggested to occur due to the effect of the triplet excited state on the nuclear configuration of the singlet excited state after RISC. Fig. 4d also shows that LPs are the source of the delayed signal in microcavity measurements.

The above results seem to indicate a negligible influence of the LP mode on the conversion of dark triplet excitons into luminescent ones. To explain this observation, we rely on a theoretical model based on the variational polaron transformation that accounts for the emergent chemical dynamics upon strong coupling[22,37]. The relevant parameters that come into play are $\lambda_S = 0.33$ eV, $\lambda_T = 0.1$ eV, and the spin-orbit coupling matrix element, $V_{ST}$, which mediates the transition between singlets and triplets. For 3DPA3CN, $\lambda_T$ has been previously estimated from $\Delta E_{st}$ and an Arrhenius plot of the RISC rate as a function of temperature[33]. Even though we do not precisely know $V_{ST}$ for this material, an approximate value is sufficient for our purposes, and we therefore consider a spin-orbit coupling interaction equal to $2 \times 10^{-2}$ meV, a characteristic value for TADF molecules[38–40]. Assuming that the transfer rate between singlet and triplet states is mediated by low-frequency modes (which seems to be the case for this particular molecule in view of the absence of vibronic progressions in the absorption spectrum) that are sufficiently fast compared to the spin-orbit coupling timescale, the population transfer between the triplet and singlet state in the bare molecules can be described in terms of a Marcus rate equation (see Fig. 5a). Theoretical analysis shows that in the strong coupling regime, the dynamics from the (dark) triplet states to the polariton modes and dark singlet states also follow an effective Marcus description[37] (see Methods), which can explain why the polariton RISC process (with rate $k_{T \to LP}$, from any triplet to the LP) is unable to compete with the bare RISC (with rate $k_{T \to DS}$, from any triplet to the dark singlet states) in our present situation.

First, let us assume that the Rabi splittings are larger than the highest frequency modes that are coupled to the exciton states. Under these circumstances, the exchange of energy between photon and singlet exciton states is much faster than the coupling of the latter to the vibrational bath, upon which so-called polaron decoupling ensues, where the LP nuclear configuration remains the same as that of the singlet ground state[21,41]. This nuclear rearrangement results in a significantly increased reorganization energy $\lambda_{T,LP}$ for the polariton RISC process when compared to $\lambda_T$, with a concomitant increase in the activation energy. This activation energy can in principle be suppressed by decreasing the energy of the LP with increased Rabi splitting, yielding an exponential increase in $k_{T\rightarrow LP}$; in fact, for the molecular parameters and light-matter couplings in the experiments above, we believe this is the case, namely, that we are operating under the normal regime of Marcus theory (larger Rabi splittings would be needed to access the inverted regime, see Fig. 5b). Yet, even if this activation energy is fully suppressed, the maximum rate is bounded at light-matter resonance by $k_{T\rightarrow LP} \leq \frac{|V_{ST}|^2}{2\hbar N_{eff}} \sqrt{\frac{\pi}{\lambda_{T,LP} k_B T}}$, where $N_{eff} = N/N_{ph}$ is the effective number of molecules coupled per cavity mode and $\hbar = h/2\pi$, with $h$ being Planck's constant. The factor of $\frac{1}{2N_{eff}}$ in this upper bound arises from the delocalization of the polariton across $N_{eff}$ molecules: half of the polariton has molecular character, and out of this fraction, only $\frac{1}{N_{eff}}$ corresponds to a singlet that can undergo RISC with a given triplet (spin-orbit coupling interactions are local). If we assume $N_{eff} = 4 \times 10^6$ (see Methods) for the MC Neat sample[42], we obtain a room-temperature $k_{T\rightarrow LP} \leq 0.1\ s^{-1}$, a value that cannot compete with the bare $k_{RISC}$ to the dark states (see Fig. 5c).

**Concluding Remarks.**

We have conclusively shown that the RISC rate in a TADF material remains invariant under the strong light-matter regime even under energy-inversion of the LP with respect to the triplet energy. Since the RISC process is thermally activated, the corresponding activation barrier to the LP can in principle be decreased by increasing the Rabi splitting. However, the large ratio $N_{eff}$ between the density of states of the dark triplet states and the polaritonic ones renders an improvement over the rate in the bare molecule unfeasible under normal conditions.

Two possible avenues to observe polariton-assisted harvesting of triplet excitons can be conjectured based on this observations. First, molecules in which the potential energy surfaces of the singlet and triplet are in the so-called inverted Marcus regime, are the best candidates to be benefit from the formation of polaritons. The reason is that the transition from the inverted to the normal Marcus regime will result in an exponential increase in the RISC rate with increased Rabi splittings, which may circumvent the large $N_{eff}$ problem. Second, polaritonic states where $N_{eff} \approx 100$ due to extreme mode volume confinement in nanoparticles[43–46] could further improve the aforementioned constraints, leading to $k_{T \to LP} \sim 10^4 \ s^{-1}$ and thus, surpassing bare RISC rates. Future work will focus on exploring these avenues.

**Methods.**

**Sample Preparation.** The devices were fabricated on glass substrates using thermal evaporation at a base pressure $< 10^{-7}$ Torr (EvoVac, Angstrom Engineering). Prior to the deposition of the films, the substrates were cleaned and exposed to a UV-ozone treatment. 3DPA3CN was purchased and used as received from Lumtec.

**Characterization**. The refractive index and the layer thicknesses were obtained using ellipsometry (J. A. Woollam Co., RC2 D+NIR). Angle resolved reflectivity measurements were performed using a Photospectrometer (Carry 7000). The angle resolved PL measurements were performed using a fiber coupled spectrometer (Ocean Optics Flame) with a 405 nm diode Laser excitation (Thorlabs CPS405). Quantum efficiency measurements were performed using an integrated sphere (Labsphere) on samples that were vacuumed in a transparent plastic to avoid exposure to oxygen.

Transient photoluminescence characteristics were measured under vacuum ($< 2 \cdot 10^{-4}$ Torr) using a Streak Camera (Hamamatsu C10910) coupled to a Spectrometer (Princeton Instruments SP-2300). The PL signal was collected as 0 degrees using an optical fiber. For delayed lifetime measurments, the samples were excited by a Nd:YAG Laser and OPO (440 nm, 10 Hz repetition rate, 8 ns pulse duration). For prompt lifetime measurements, the samples were excited by a supercontinuum laser (Fianium WhiteLaser UV, 0.1 MHz repetition rate, ~50 ps pulse duration, 405 nm using optical filters). The mean lifetimes were calculated as $\tau_m = \frac{\sum_i a_i \tau_i^2}{\sum_i a_i \tau_i}$ from multi exponential fits to the data (two terms for the prompt lifetime and three for the delayed), where $a_i$ and $\tau_i$ are each exponent weight and lifetime respectively. The ratios, $\Phi_d/\Phi_p$, were obtained using integration over the PL decays of delayed and prompt components. Low-temperature measurements were conducted in vacuum using a closed-cycle optical cryostat.

**Theoretical Modeling**.

**Ultrastrong coupling.** The experimental Rabi splitting and the detuning were extracted for TE or TM modes by performing a least-square fit of the reflectivity dips to the following linear equation[34,47,48]

$$\begin{pmatrix} E_{ph}(q)+2D & \Omega/2 & 2D & \Omega/2 \\ \Omega/2 & E_x & \Omega/2 & 0 \\ -2D & -\Omega/2 & -E_{ph}(q)-2D & -\Omega/2 \\ -\Omega/2 & 0 & -\Omega/2 & -E_x \end{pmatrix} \begin{pmatrix} w_{j,q} \\ x_{j,q} \\ y_{j,q} \\ z_{j,q} \end{pmatrix} = E_{j,q} \begin{pmatrix} w_{j,q} \\ x_{j,q} \\ y_{j,q} \\ z_{j,q} \end{pmatrix} \quad (3)$$

Equation 3 is a solution of Hopfield Hamiltonian and the eigenvalues correspond to the polariton energies $\pm E_{LP,q}$ and $\pm E_{UP,q}$. The in-plane wavevector is $q$, $j \in \{LP, UP\}$ and $D = \Omega^2/4E_x$. We used a photonic cavity dispersion with an effective refractive index $n_{eff}$,

$$E_{ph}(q) = \sqrt{(\frac{hcq}{n_{eff}})^2 + E_c^2}, \quad (4)$$

with $q = \frac{\omega}{c}\sin\theta$, where $\theta$ is the angle of incidence.

**Rate calculations.** We used a variational polaron transformation of the Hamiltonian (describing electronic states, photonic mode, and vibrational degrees of freedom) to account for the dissipative dynamics that emerge under strong light-matter coupling. This approach introduces an optimal energy partitioning between a vibrationally-dressed polaritonic system (featuring renormalized energies of the triplet, singlet, and photonic states, as well as their couplings), and a small perturbation that contains residual couplings mediated by the vibrational bath[37]. The latter perturbation is dealt with in the Markovian approximation using standard open-quantum systems techniques. For the parameters $\lambda_S$, $\lambda_T$, and $V_{ST}$ assumed for 3DPA3CN, the rate between any triplet and the LP in the strong coupling regime is well described by a Marcus equation of the form

$$k_{T \to LP} = \frac{|x_{j,q=0}|^2}{N_{eff}} \times \frac{|V_{ST}|^2}{\hbar} \sqrt{\frac{\pi}{\lambda_{T,LP} k_B T}} \exp\left[-\frac{(\Delta E_{ST} - \Omega/2 + \lambda_{T,LP})^2}{4\lambda_{T,LP} k_B T}\right] \quad (5)$$

where $\lambda_{T,LP} = \left(\sqrt{\lambda_S} + \sqrt{\lambda_T}\right)^2$, is the reorganization energy between the triplet and the (polaron decoupled) lower polariton state. Notice that the rate above contains a prefactor $\frac{|x_{j,q=0}|^2}{N_{eff}}$ which equals the probability of finding the singlet in the LP state which can accept population from a given triplet. We estimate $N_{eff}$ is based on the ratio of the molecular density of states to the cavity photon counterpart.[42,49] The latter is given by $\rho_{ph} = \frac{n_{eff}^2}{4\pi}\left[E_{ph}^2(q_{cut}) - E_c^2\right]$, where $E_{ph}(q_{cut})$ is the energy cutoff that we chose based on the range of angles which exhibit the largest PL signals for MC Neat (see Fig. S1). By selecting $\theta = 60^0$ for the latter sample, we have $E_{ph}(q_{cut}) = 2.6$ eV and $E_c = 2.3$ eV, which yields $\rho_{ph} = 12 \, \mu m^{-2}$. The density of molecular states can be estimated from the number density of the molecular emitters, approximately $7 \times 10^{20}$ cm$^{-3}$, which for a thickness of 70 nm corresponds to $\rho_{mol} = 4.9 \times 10^7 \, \mu m^{-2}$. Therefore, we obtain $N_{eff} = \frac{\rho_{mol}}{\rho_{ph}} = 4 \times 10^6$.

AUTHOR INFORMATION

*Corresponding authors: E.E. elad.eizner@polymtl.ca, S. K-C. s.kena-cohen@polymtl.ca.

Notes

The authors declare no competing financial interest.

ACKNOWLEDGMENT

The authors acknowledge funding from the NSERC Discovery Grant Program (RGPIN-2014-06129) and the Canada Research Chairs program. L.A.M.M is grateful for the support of the UC-Mexus CONACyT scholarship for doctoral studies. J.Y.Z. and L.A.M.M. acknowledge support from NSF EAGER CHE-1836599.

.

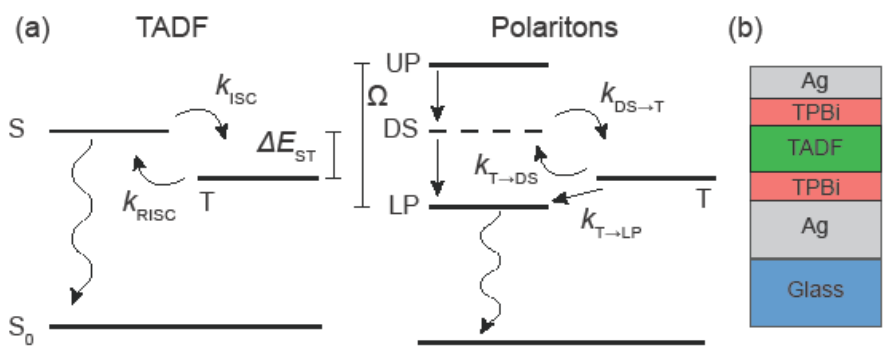

**Figure 1.** (a) Electronic energy levels and rate constants for the TADF material and polaritons. (b) Microcavity structure consisting of Ag bottom mirror (100 nm), Ag top mirror (30 nm), TBPi buffer layers (10 nm each) and 3DPA3CN.

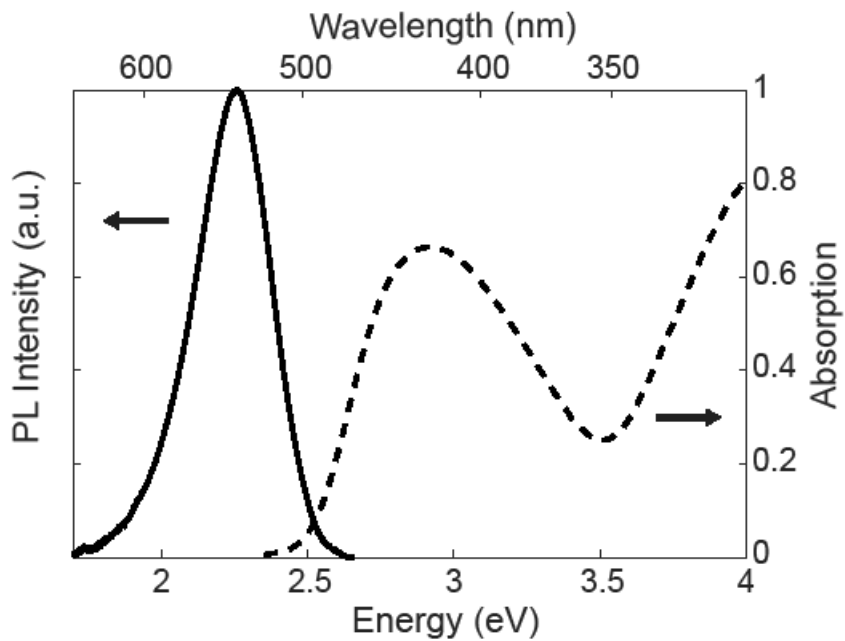

**Figure 2.** Absorption (dashed line) and PL (solid line) spectra of 3DPA3CN neat film.

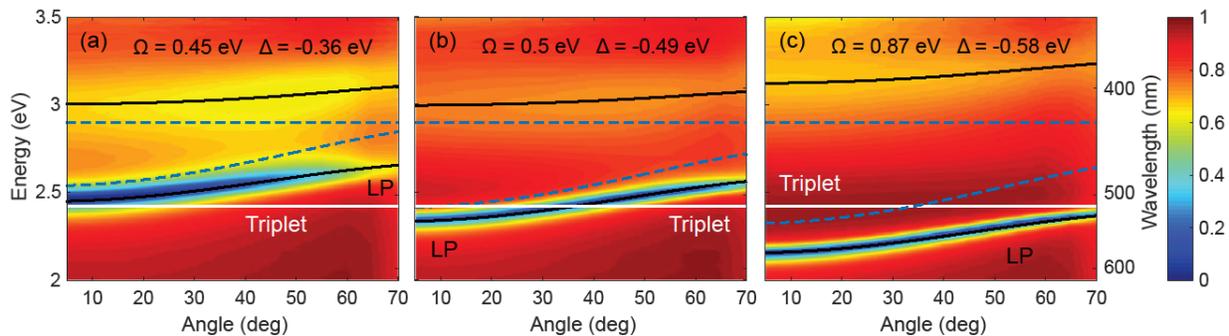

**Figure 3.** Angle resolved reflectivity measured using TE polarized light for (a) MC 1, (b) MC 2, and (c) MC Neat (as defined in the text). The solid black lines show the least square fit to the Hopfield Hamiltonian (see Methods). The dashed blue lines are the energies corresponding to the uncoupled photonic mode and the exciton singlet absorption. The solid white line is the triplet energy, which is $0.1\text{eV}$[33] lower in energy than the bottom of the singlet band.

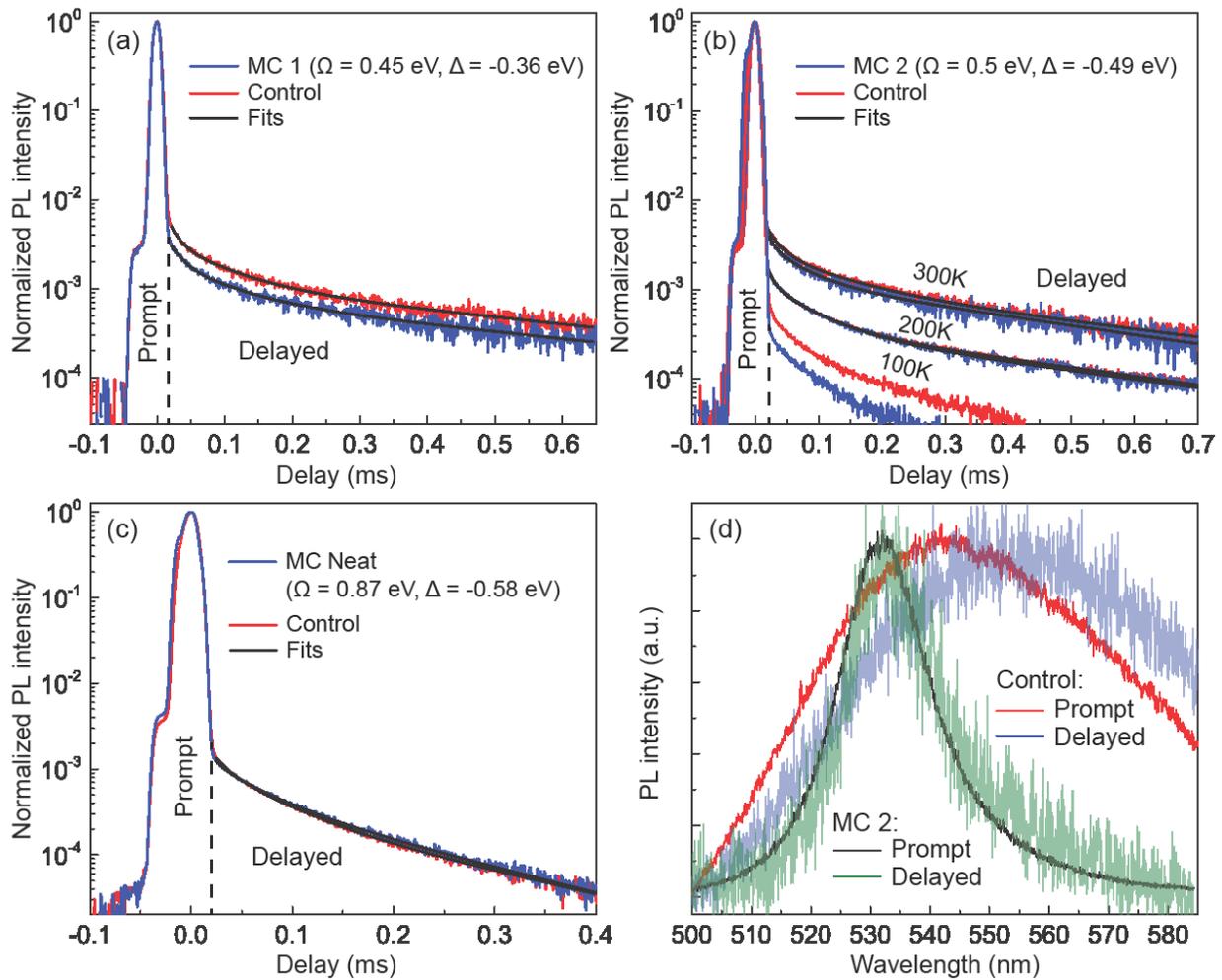

**Figure 4.** Transient PL decays for the LP (blue line) and control film singlet (red line) at a collection angle of zero degrees. (a) MC 1, (b) MC 2 and (c) MC Neat (as defined in the text). The black lines are multi-exponential fits to the delayed signal decay. (d) Delayed and prompt PL spectra for MC 2 and control at room temperature.

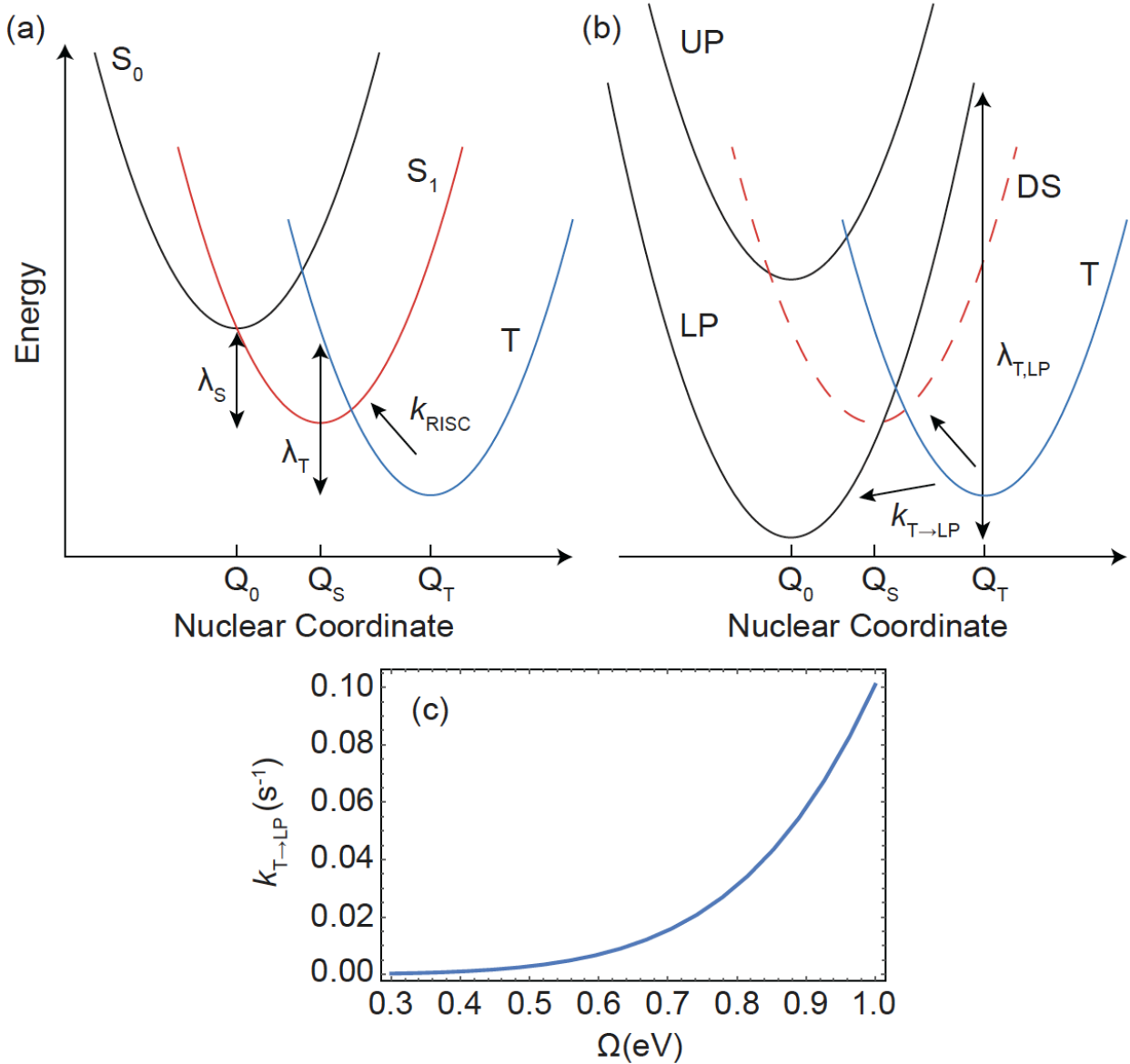

**Figure 5.** Schematic representation of the nuclear potential energy surfaces and population dynamics of the triplet and singlet electronic excited states. (a) In the cavity-free scenario, the population accumulated in the triplet electronic manifold needs to overcome a thermal energy barrier to reach the fluorescent singlet electronic state. (b) Upon strong coupling, the energy barriers are different for the dark states and the polariton ones. Transfer to the LP features a smaller energy barrier compared to the bare case. However, the delocalization of the LP across $N_{eff} \approx 10^6$ molecules dramatically dilutes the electrostatic coupling between a triplet at a given molecule and the latter, making the polariton RISC very slow and not competitive with the bare one. This is shown in (c), where we calculate polariton RISC rates for the range of Rabi splittings $\Omega$ explored in our experiments (we assume a detuning of $\Delta = -0.58 \, eV$ as in the MC Neat sample at $q = 0$); the computed rates are still lower than the bare RISC rate for $N_{eff} = 4 \times 10^6$.

| Sample | $k_p \cdot 10^8$ [1/s] | $k_d \cdot 10^3$ [1/s] | $\Phi_d/\Phi_p$ | $\Phi$ | $k_{ISC} \cdot 10^7$ [1/s] | $k_{RISC} \cdot 10^3$ [1/s] |
|---|---|---|---|---|---|---|
| MC 1 | 2.2 ± 0.1 | 2.5 ± 0.3 | 4.0% | 7.5 ± 3% | `1.4 ± 0.6 | 1.5 ± 0.6 |
| Control 1 | 1.8 ± 0.1 | 2.6 ± 0.2 | 6.3% | 98 ± 3% | 1.4 ± 0.6 | 2.1 ± 0.8 |
| MC 2 | 2.2 ± 0.1 | 3.2 ± 0.2 | 4.5% | 32 ± 3% | 1.4 ± 0.4 | 2.3 ± 0.7 |
| Control 2 | 1.4 ± 0.1 | 2.8 ± 0.3 | 5.4% | 95 ± 3% | 1.4 ± 0.4 | 1.5 ± 0.5 |
| MC 2 (200 K) | 2.2 ± 0.1 | 3.4 ± 0.1 | 1.2% | 31 ± 3% | 1.3 ± 0.5 | 0.7 ± 0.3 |
| Control 2 (200 K) | 1.6 ± 0.1 | 3.3 ± 0.3 | 1.2% | 93 ± 3% | 1.3 ± 0.5 | 0.5 ± 0.2 |
| MC 2 (100 K) | 2.3 ± 0.1 | – | 0.2% | 32 ± 3% | 0.9 ± 0.5 | – |
| Control 2 (100 K) | 1.7 ± 0.1 | – | 0.3% | 95 ± 3% | 0.9 ± 0.5 | – |
| MC Neat | 3.0 ± 0.1 | 10.9 ± 1.8 | 0.7% | 20 ± 3% | 2.8 ± 0.6 | 0.8 ± 0.2 |
| Control Neat | 1.9 ± 0.1 | 11.3 ± 0.9 | 0.7% | 86 ± 3% | 2.8 ± 0.6 | 0.6 ± 0.1 |

**Table 1.** Photophysical data of the studied microcavities and the corresponding control films. MC 1 ($\Omega = 0.45$ eV, $\Delta = -0.36$ eV), MC 2 ($\Omega = 0.5$ eV, $\Delta = -0.49$ eV) and MC Neat ($\Omega = 0.87$ eV, $\Delta = -0.58$ eV). Where not specified, measurements were performed at room temperature. Notes: (i) $k_{ISC}$ for the microcavities were taken from the control values. (ii) The PLQEs for 100 K and 200 K were estimated from the measured yields: $\Phi_{PF}(200 \text{ K})/\Phi_{PF}(300 \text{ K}) = 1.02$, $\Phi_{PF}(100 \text{ K})/\Phi_{PF}(300 \text{ K}) = 1.05$ for the control, and $\Phi_{PF}(100 \text{ K})/\Phi_{PF}(300 \text{ K}) = 1.01$ for MC 2. (iii) The uncertainties of $k_d$ and $k_p$ were estimated from the fit's standard deviation errors. For the PLQEs, a constant 3% uncertainty error was assumed. (iv) Even for long integration times, the delayed signal at 100 K is too weak to reliably extract the delayed lifetime. (v) From the two data points of the RISC rates at 200 K and 300 K, using Eq. 1 we obtain activation energy, $\Delta E_a = 60$ meV.

# Supporting Information:

| Sample | Polarization | $\Omega$ [eV] | $n_{eff}$ | $E_c$ [eV] |
|--------|--------------|---------------|-----------|------------|
| MC 1   | TE           | $0.45 \pm 0.02$ | $2.08 \pm 0.11$ | $2.54 \pm 0.02$ |
| MC 1   | TM           | $0.40 \pm 0.01$ | $3.50 \pm 0.22$ | $2.54 \pm 0.02$ |
| MC 2   | TE           | $0.50 \pm 0.02$ | $2.04 \pm 0.10$ | $2.41 \pm 0.03$ |
| MC 2   | TM           | $0.42 \pm 0.02$ | $3.26 \pm 0.49$ | $2.41 \pm 0.03$ |
| MC Neat | TE          | $0.87 \pm 0.01$ | $1.97 \pm 0.06$ | $2.32 \pm 0.02$ |
| MC Neat | TM          | $0.87 \pm 0.02$ | $2.86 \pm 0.39$ | $2.32 \pm 0.02$ |

**Table S1.** Parameters obtained from a least square fit of the reflectivity data to the Hopfield Hamiltonian.

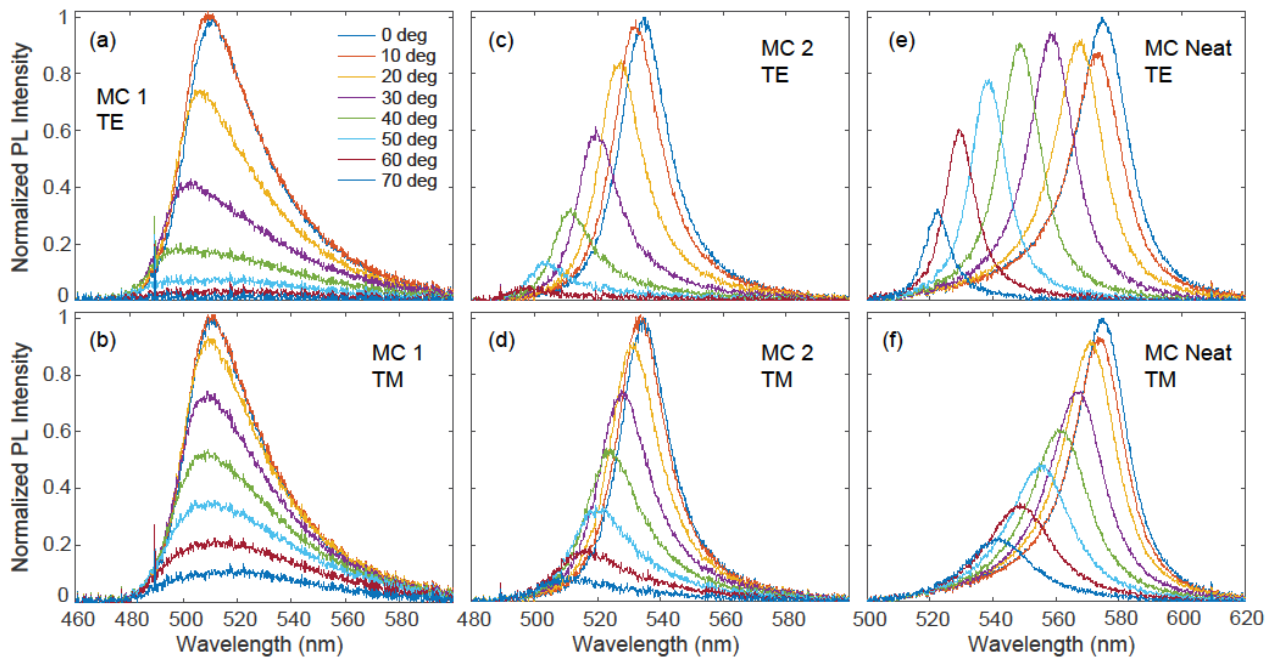

**Figure S1.** Angle resolved PL for MC 1 ($\Omega = 0.45$ eV, $\Delta = -0.36$ eV) at (a) TE and (b) TM polarization, for MC 2 ($\Omega = 0.5$ eV, $\Delta = -0.49$ eV) at (c) TE and (d) TM polarization, and for MC Neat ($\Omega = 0.87$ eV, $\Delta = -0.58$ eV) at (e) TE and (f) TM polarization.

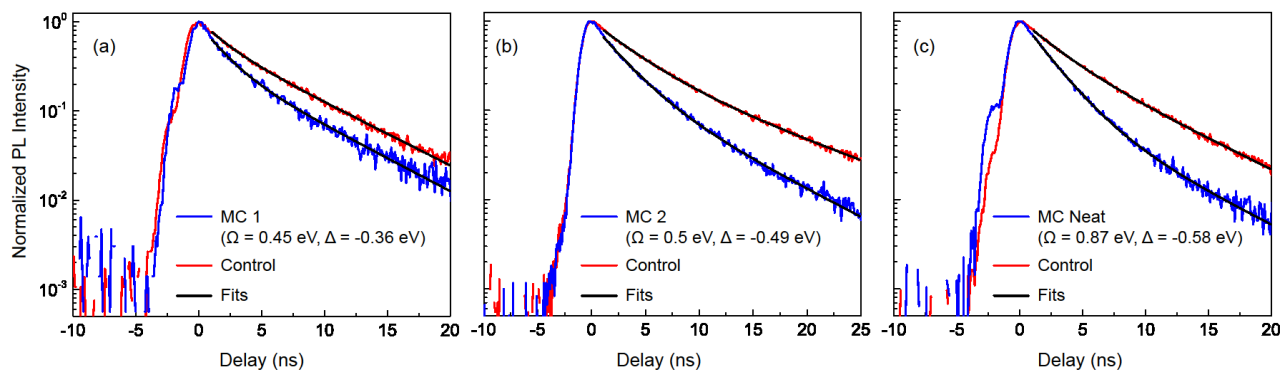

**Figure S2.** Transient prompt PL decays for the LP (blue line) and control film (red line). (a) MC 1, (b) MC 2 and (c) MC Neat. The black lines are multi-exponential fits to the prompt signal decay.